\documentclass[a4paper]{jpconf}
\begin{document}
\title{Experimental Status of Exotic Mesons and the GlueX Experiment}

\author{Daniel S. Carman}

\address{Jefferson Laboratory, Newport News, VA 23606}

\ead{carman@jlab.org}

\begin{abstract}
One of the unanswered and most fundamental questions in physics regards
the nature of the confinement mechanism of quarks and gluons in QCD. 
Exotic hybrid mesons manifest gluonic degrees of freedom and their 
spectroscopy will provide the data necessary to test assumptions in 
lattice QCD and the specific phenomenology leading to confinement. Within 
the past two decades a number of experiments have put forth tantalizing 
evidence for the existence of exotic hybrid mesons in the mass range below 
2~GeV.  This talk represents an overview of the available data and what 
has been learned.  In looking toward the future, the GlueX experiment at 
Jefferson Laboratory represents a new initiative that will perform detailed 
spectroscopy of the light-quark meson spectrum.  This experiment and its 
capabilities will be reviewed.
\end{abstract}

\section{Motivation for the Study of Hybrid Mesons}

High energy experiments have provided clear evidence for significant
contributions of gluons to hadronic structure.  Evidence for gluons
has been found in jet measurements and deep inelastic scattering.
However, descriptions of gluonic degrees of freedom in the low energy
regime of soft gluons are still unavailable. This description is
necessary to better understand the detailed nature of confinement.
The nature of this mechanism is one of the great mysteries of modern
physics, and in order to shed light on this phenomenon, we must better
understand the nature of the gluon and its role in the hadronic
spectrum.  Confinement within the theory of strongly interacting matter,
Quantum Chromodynamics (QCD), arises from the postulate that gluons
can interact among themselves and give rise to detectable signatures
within the hadronic spectrum.  These signatures are expected within
hadrons known as hybrids, where the gluonic degree of freedom is excited
and can provide for a more detailed understanding of the confinement 
mechanism in QCD.

\begin{figure}[htbp]
\vspace{6.2cm}
\includegraphics{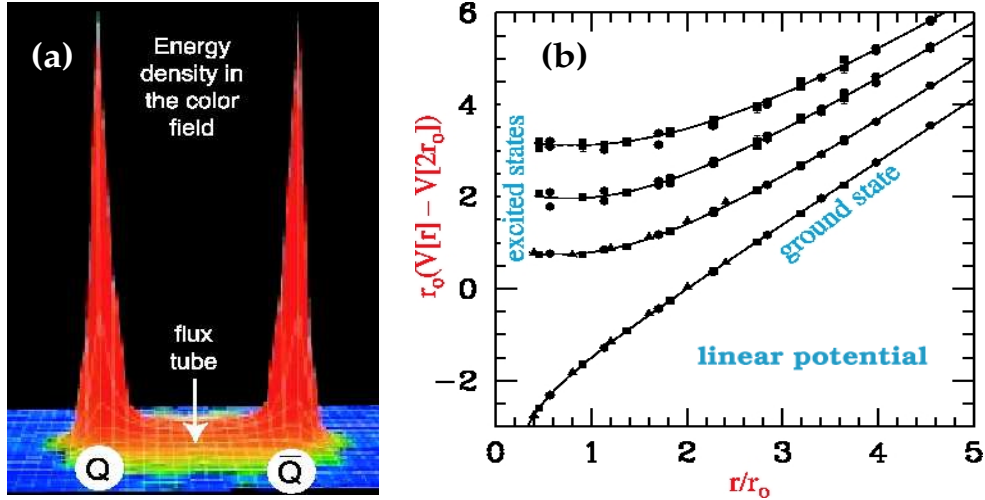}
\caption{(a). A lattice QCD calculation of the energy density in the color 
field between a quark and an antiquark~\cite{bali_fig}.  This density peaks 
at the locations of the $q$ and $\bar{q}$ and is confined to a flux-tube 
stretching between the pair. (b). Corresponding lattice calculation of 
the potential between the quarks showing the ground and low-lying excited
states~\cite{bali}.  These calculations are for heavy quarks in the quenched
approximation.}
\label{lqcd}
\end{figure}

Gluonic mesons represent a $q\bar{q}$ system in which the gluonic flux-tube
contributes directly to the quantum numbers of the state.  In terms of
the constituent quark model, the quantum numbers of the meson are determined 
solely from the quark and antiquark.  However, QCD indicates that this simple 
picture is incomplete.  Lattice QCD calculations predict that hybrid states 
with the flux-tube carrying angular momentum should exist, as well
as purely gluonic states (called glueballs).  Modern lattice calculations for
mesons show that indeed a string-like chromoelectric flux-tube forms between 
distant static quark charges as shown in Fig.~\ref{lqcd}a.  The 
non-perturbative nature of the flux-tube leads to the confinement of the 
quarks and to the well-known linear inter-quark potential from heavy-quark
confinement with $dV/dr \sim 1$~GeV/fm (see Fig.~\ref{lqcd}b).  These 
calculations predict that the lowest lying hybrid meson states are roughly 
1~GeV more massive than the conventional meson states.  This provides a 
reference point for the mass range to which experiments must be sensitive.

\section{Hybrid Meson Properties}

Within the standard non-relativistic constituent quark model, conventional 
mesons are made up from $q\bar{q}$ pairs with the spin 1/2 quarks coupled to 
a total spin $S$=0 or 1.  These pairs are then coupled with units of orbital 
angular momentum $L$, and a possible radial excitation.  The relevant quantum 
numbers to describe these states for a given principal quantum number $n$ are 
$J^{PC}$, where $\vec{J}=\vec{L}+\vec{S}$ represents the total angular 
momentum, $P=(-1)^{L+1}$ represents the intrinsic parity, and $C=(-1)^{L+S}$ 
represents the $C$-parity of the state.

The light-quark mesons are built up from $u$, $d$, and $s$ quarks and their
antiquarks.  For each value of $S$, $L$, and $n$, a nonet of mesons is 
expected for each value of $J^{PC}$.  The lowest possible mass states for 
these conventional mesons (i.e. for $L$=0, $n$=1) are then $J^{PC} = 0^{-+}$ 
for $S$=0 (corresponding to the $\pi$, $\eta$, $\eta'$, and $K$ mesons) and 
$J^{PC}=1^{- -}$ for $S$=1 (corresponding to the $\rho$, $\omega$, $\phi$, 
and $K^*$ mesons).  The lowest lying $S$=0 mesons have masses of about 500~MeV 
and the lowest lying $S$=1 mesons have masses of about 800~MeV.  Nonets of 
higher mass mesons for a given principal quantum number are then built up by 
adding in units of $L$.  Conventional mesons correspond to states with the 
flux-tube in its ground state, and as such, the gluonic degree of freedom 
does not contribute.

The hybrid meson quantum numbers can be predicted within the flux-tube
model~\cite{Is85}.  In its ground state, the flux-tube carries no angular 
momentum.  The lowest excitation is an $L$=1 rotation which contains two 
degenerate states (corresponding to clockwise and counterclockwise rotations).
Linear combinations of these states give rise to quantum numbers of 
$J^{PC} = 1^{- +}$ or $1^{+ -}$ for the flux-tube.  Adding these quantum 
numbers to those for the $q\bar{q}$ pair gives the possible $J^{PC}$ for 
the hybrid mesons, which also are expected to have a nonet of states for 
each $J^{PC}$.  For $S=L=0$, the possible quantum numbers are 
$J^{PC}= 1^{- -}$ and $1^{+ +}$.  Note that exotic hybrid mesons are 
{\em not} generated when $S$=0, as these quantum numbers are possible for 
conventional $q\bar{q}$ states such as $\rho$, $\omega$, and $\phi$.  An 
important consideration in the study of these states is that non-exotic 
hybrids may mix with conventional $q\bar{q}$ states making clear experimental 
identification difficult.  Establishing the hybrid nonets will depend on 
starting with nonets whose quantum numbers are exotic to which ordinary 
$q\bar{q}$ states cannot couple.

For $S$=1, $L$=0 states the possible quantum numbers are $J^{PC} = 0^{+ -}$,
$0^{- +}$, $1^{- +}$, $1^{+ -}$, $2^{+ -}$, $2^{- +}$.  Of these states,
those with $J^{PC} = 0^{+ -}$, $1^{- +}$, and $2^{+ -}$ are manifestly
exotic, i.e. they are not allowed for ordinary $q\bar{q}$ states.  The 
existence of exotic hybrid mesons has been predicted for more than 3-decades.  
Almost all QCD models predict a $J^{PC}=1^{- +}$ hybrid with a mass at or 
below 2~GeV~\cite{Is85,barnes,ChSh,BaDy}, with favored widths in the range 
$\Gamma \sim 40 \rightarrow 100$~MeV~\cite{lattdecay}. However, the decay 
modes are uncertain. In the flux-tube model, the gluonic excitation does not 
transfer its spin to the relative orbital angular momentum of the final state 
mesons, and hybrid decay to two mesons with quarks in $S$-waves does not 
occur~\cite{IsgKok}. However, other models predict such a decay, through 
effects such as spurious bag CM motion~\cite{ChSh}, or through the sequential 
decay of the exotic hybrid into a non-exotic hybrid and then into a 
conventional meson via mixing~\cite{tan}. The non-exotic hybrid is mixed with 
a conventional meson which appears in the final state. 

\begin{figure}[htbp]
\vspace{6.3cm}
\includegraphics{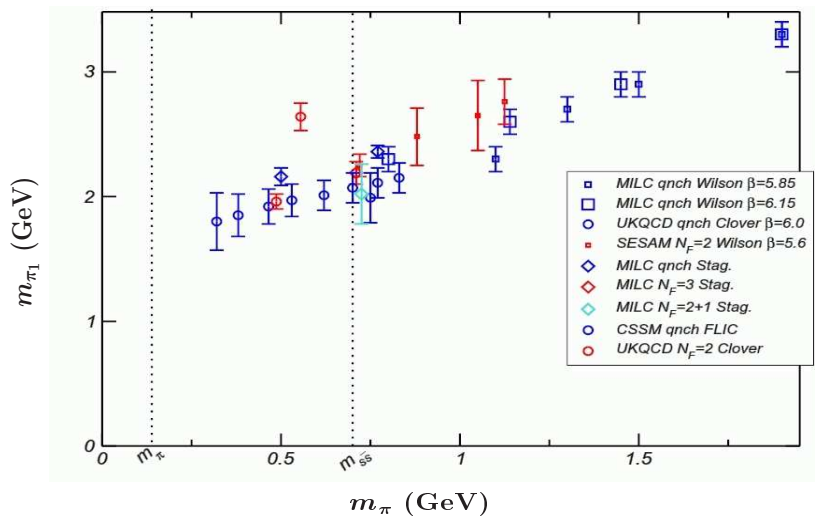}
\caption{A survey of current LQCD results for the lightest $J^{PC}=1^{- +}$ 
exotic plotting $m_{\pi_1}$ vs. $m_\pi$.  The open and closed symbols denote 
dynamical and quenched simulations, respectively.}
\label{lqcd_proc}
\end{figure}

A summary of current LQCD calculations for the lowest lying non-strange 
$J^{PC}=1^{- +}$ exotic meson is provided in Fig.~\ref{lqcd_proc}
\cite{michael,dudek}.  These quantum numbers should correspond to the 
lightest nonet. These calculations are quenched or partially quenched at the 
current time with some employing improved lattice actions, and they indicate 
that the rough mass scale to search in is about 1.5 - 2.0~GeV.

\boldmath
\section{Experimental Status of $J^{PC}$=1$^{-+}$ Hybrids}
\unboldmath

While the current results from lattice QCD indicate that the lightest 
exotic meson nonet has quantum numbers $J^{PC}$=1$^{-+}$ and its lightest 
member, the $\pi_1$, has a mass in the range from 1.5 to 2.0~GeV, the 
current experimental evidence is much less clear.  During the past two
decades, a number of different experiments have provided tantalizing evidence 
for the $1^{-+}$ exotics $\pi_1$(1400), $\pi_1$(1600), and $\pi_1$(2000).  
If each of these states was verified, this would result in an 
overpopulation of the $1^{-+}$ hybrid nonet where there should be only one 
$\pi_1$ state.  In this section I provide a brief overview on each of the 
reported $1^{-+}$ exotic candidates.

Beyond issues associated with the statistics in each of these
analyses, the different experiments are hampered by a number of analysis 
issues.  Perhaps the most important arises due to leakage effects in the 
amplitude analysis.  While the implementation of a partial wave analysis 
is in principle straightforward, there are difficulties that arise due to 
the detector system employed, as well as ambiguities within the partial
wave analysis (PWA) framework itself.  Effects such as detector acceptance 
and resolution can conspire to allow strength from a dominant partial wave 
to appear as strength in a weaker wave.  Another important issue involves 
the model-dependent assumptions made within the PWA framework itself.  In 
PWA, simplifying assumptions are used in order to make the fitting model 
more tractable, such as in calculating decay amplitudes via an isobar model, 
and the absolute effects of these assumptions are not fully known.

\subsection{Evidence for the $\pi_1$(1400)}
\label{pi1400}

The $\pi_1$(1400) was first reported by the GAMS group at CERN~\cite{gams}.  
It was seen in the $\pi^- p \to \eta \pi^0 n$ channel at $p_\pi$=100~GeV.  
A partial wave analysis of these data showed a clear $a_2$(1320) $D$-wave, 
and a narrow enhancement in the unnatural parity exchange $P_0$-wave 
at a mass of $\sim$1.4~GeV. The natural parity exchange $P_+$-wave was 
observed structureless.  However, these conclusions were refuted by some 
of the original authors who pointed out that the $\eta\pi^0$ PWA suffered 
from an eight-fold ambiguity which was not properly accounted for in the 
analysis~\cite{inna95,prok95}.  The VES Collaboration at IHEP~\cite{ves1}, 
which studied the reaction $\pi^- N \to \pi^- \eta N$ with $p_\pi$=37~GeV, 
reported a small but statistically significant broad enhancement in the 
natural parity exchange $P_+$-wave at about 1.4~GeV.  The authors did not 
identify the $P_+$ enhancement as a resonance.  Experiment 
E179 at KEK studied the decay angular distributions in the 
$\pi^- p \to \eta \pi^- p$ reaction at $p_\pi$=6.3~GeV.  They observed an 
enhancement around 1.3~GeV in the $P_+$-wave but were unable to establish a 
resonant nature. They noted the phase of the $P_+$-wave relative to the 
$D_+$ $a_2(1320)$ wave showed no distinct variation with mass~\cite{kek}. 

The first resonant claim of the $\pi_1$(1400) was provided by the E852 
Collaboration at BNL in the reactions $\pi^- p \to \eta \pi^0 n$ and 
$\pi^- p \to \eta \pi^- p$ at $p_\pi$=18~GeV~\cite{thompson}. In this 
analysis the resonant nature of the $P_+$-wave arises from a strong 
interference with the $D_+$ $a_2(1320)$ wave.  The phase difference between 
these waves exhibited a phase motion not attributable solely to the 
$a_2(1320)$.   Results from the Crystal Barrel experiment at CERN for the 
reactions $\bar{p} p \to \eta \pi^0 \pi^0$ and 
$\bar{p} n \to \eta \pi^- \pi^0$ at $p_{\bar{p}}$=200~MeV~\cite{cbarrel}, 
needed to include a $\pi_1$(1400) state in addition to conventional mesons 
to fit their data.

In the PDG listings~\cite{pdg}, the mass of the $\pi_1$(1400) is 
$M$=1376$\pm$17~MeV and its width $\Gamma$=300$\pm$40~MeV with observed 
decays to $\eta \pi^0$ and $\eta \pi^-$.  However the current experimental 
evidence gives rise to a number of controversial issues.  The $\pi_1$(1400) 
is significantly lighter than theoretical expectations, and its only observed 
decay mode into $\eta \pi$ is not expected (or strongly disfavored) for a 
gluonic hybrid. It has been suggested that the $\pi_1$(1400) could represent 
a meson-meson molecule. An alternative suggestion is that the $\pi_1$(1400) 
could actually be a threshold effect of a higher mass $\pi_1$ resonance due 
to the opening of more favorable decay channels. Other recent work suggests 
that the exotic $P$-wave signature for the $\pi_1$(1400) may actually arise 
from dynamical non-resonant scattering, similar to $S$-wave $\pi \pi$ 
scattering at low energy~\cite{adam}.

\begin{figure}[htbp]
\vspace{8.8cm} 
\includegraphics{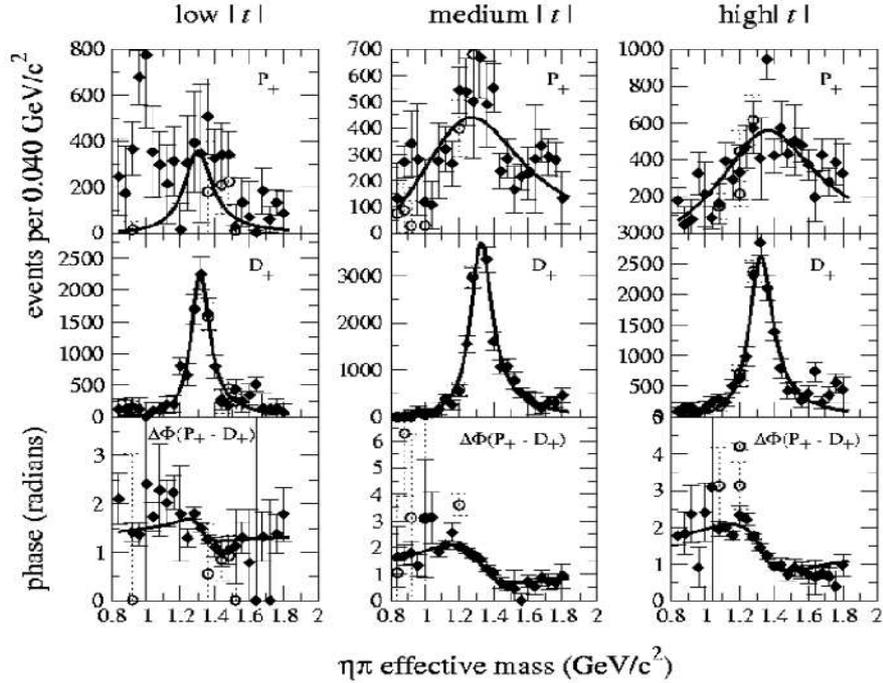}
\caption{PWA solution for the $P_+$ (top) and $D_+$ (middle) waves,
and the phase difference (bottom) for the three $t$ bins used in the
analysis as a function of the $\eta \pi^0$ effective mass~\cite{dzierba03}.}
\label{pi1400_c}
\end{figure}

A recent analysis of E852 data by Dzierba {\it et al.} focussed on an 
amplitude analysis of the $\eta \pi^0$ mode in three different $t$ bins 
(see Fig.~\ref{pi1400_c})~\cite{dzierba03}.  This analysis concluded 
that no consistent $P$-wave resonant parameters can describe the data for 
the $\pi_1$(1400), while the resonant parameters obtained for the $a_2$(1320) 
phase reference state are consistent for the different $t$ bins with the PDG 
values.  However the question remains as to what causes the peaking in the 
$P$-wave intensity distributions, and if it is due to a non-resonant source, 
what explains the phase motion with respect to the $a_2$(1320)?

Another interesting fact is that while the exotic wave is a few percent of 
the dominant $a_2$ wave in the E852 data, it is of comparable strength to 
the $a_2$ in the Crystal Barrel results.  However, each of the existing 
data sets in these hadroproduction experiments is hampered by relatively 
low statistics.

\subsection{Evidence for the $\pi_1$(1600)}

A second $J^{PC}$ = 1$^{-+}$ exotic meson at 1.6~GeV has been claimed
by the BNL E852 Collaboration in the $\eta' \pi$, $\rho \pi$, $b_1 \pi$,
and $f_1 \pi$ final states in $\pi^- p$ reactions at $p_\pi$=18~GeV
\cite{e852b,ivanov,f1ref,b1ref}. This signal first appeared as an 
enhancement in the $P_+$-wave in an early $\eta' \pi$ measurement by the 
VES Collaboration~\cite{ves1}.  Additional VES measurements followed with 
confirmation of the $\pi_1$(1600) decaying into $b_1 \pi$, $\eta' \pi$, 
and $\rho \pi$~\cite{ves2}.  This work provided measurements of the 
relative branching ratios into $b_1 \pi$, $\eta' \pi$, and $\rho \pi$ of 
1.0:1.0:1.6.  These predictions are highly at odds with predictions of the 
flux-tube model.  Thus either these three modes are not all due to a hybrid 
meson or there is a problem with the amplitude analysis or the flux-tube 
model.

\begin{figure}[htbp]
\vspace{9.8cm} 
\includegraphics{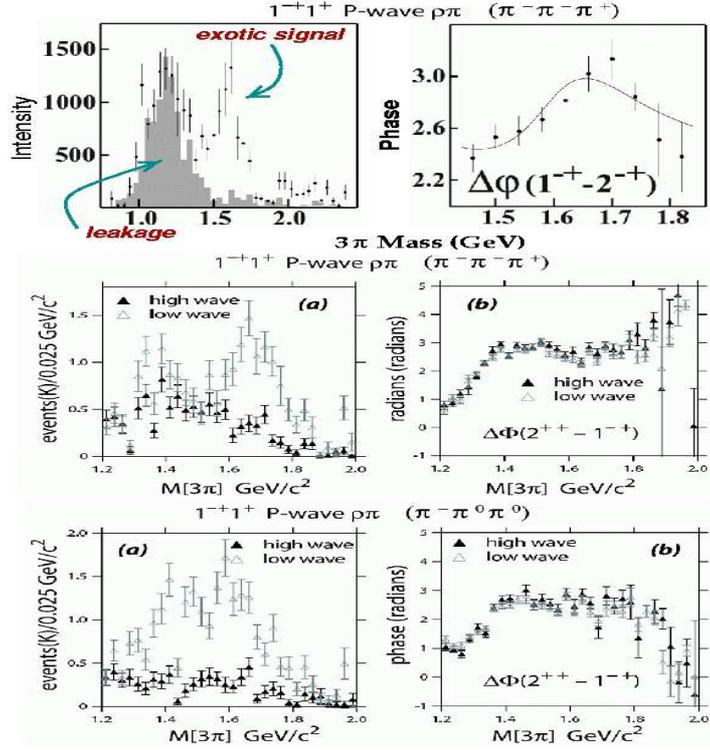}
\caption{PWA results from analysis of E852 data for the $\pi^- \pi^- \pi^+$ 
channel (top and middle rows) and for the $\pi^- \pi^0 \pi^0$ channel (bottom 
row) as a function of the $3\pi$ effective mass.  The upper row of plots is 
from Ref.~\cite{e852b} and the bottom two rows are from Ref.~\cite{dzierba06}.}
\label{pi1600_1_ss}
\end{figure}

The evidence for the $\pi_1$(1600) from its decay into $\rho \pi$ is 
particularly controversial.  The analysis results for the $\rho \pi$ final 
state are shown in Fig.~\ref{pi1600_1_ss}.  The top row of this figure shows 
the $\rho \pi$ analysis from E852~\cite{e852b} from the 1994 data run based 
on 250k events.  Here an exotic $1^{-+}$ signal is reported at a mass of 
$M$=1593$\pm$8~MeV with a width $\Gamma$=168$\pm$20~MeV.  The middle (bottom)
row of Fig.~\ref{pi1600_1_ss} shows analysis of the 1995 E852 data by Dzierba 
{\it et al.}~\cite{dzierba06} for $\rho \pi$ decay to $\pi^- \pi^- \pi^+$ 
($\pi^- \pi^0 \pi^0$).  This later data run had roughly four times the 
statistics compared to the earlier run, and highlighted a strong
PWA model dependence of the shape and magnitude of the $1^{-+}$ signal.
The $1^{-+}$ intensity distribution exhibits a strong resonance-like 
distribution using a PWA model with a minimum number of partial 
waves (21 waves).  However, when a larger wave set is used (denoted
as the high wave set in Fig.~\ref{pi1600_1_ss} -- includes 35 waves),
the evidence for the $\pi_1$(1600) in the intensity distributions is
washed out.  Nonetheless, it is curious that the phase motion plots
are essentially unchanged between the two choices of wave sets.
The comparison of the two $(3 \pi)^-$ modes provides powerful cross
checks on the analysis results.  Any resonance decaying to
$(\rho \pi)^-$ should decay equally to $(\pi^- \pi^0)\pi^0$ and
$(\pi^+ \pi^-)\pi^0$, and thus appear with equal probabilities in the
two modes.

\begin{figure}[htbp]
\vspace{6.5cm} 
\includegraphics{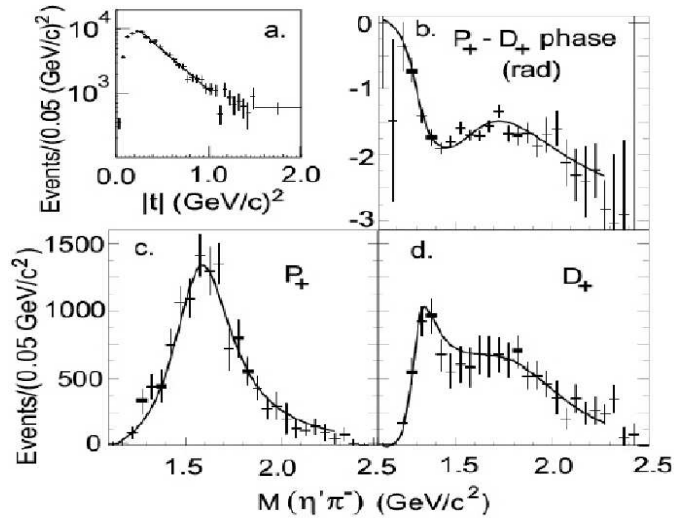}
\caption{(a) The $t$-distribution of the $\eta' \pi$ data.
(b)-(d) PWA results for the $P_+$ and $D_+$ waves as a function of the 
$\eta' \pi$ effective mass along with the results of a mass-dependent 
fit~\cite{ivanov}.}
\label{pi1600_2_ss}
\end{figure}

In the E852 analysis of $\pi^- p \to \eta' \pi$, evidence for an exotic 
$1^{-+}$ state with a mass $M$=1597$\pm$10~MeV and width
$\Gamma$=340$\pm$40~MeV is shown~\cite{ivanov} (see 
Fig.~\ref{pi1600_2_ss}).  It is interesting that the $P$-wave strength is 
the dominant signal in $\eta' \pi$ compared to $\rho \pi$. However, the 
strength in the $D$-wave used as the phase reference in the  $\eta' \pi$ 
analysis is not well understood.  So while there are some noted controversies 
associated with the $\pi_1$(1600), there clearly are some tantalizing hints 
and effects from a number of experiments that need to be further investigated.

\subsection{Evidence for the $\pi_1$(2000)}

A final candidate $J^{PC}$=1$^{-+}$ meson in the literature is the
$\pi_1$(2000) which has been claimed by the E852 Collaboration
through its decays into $f_1 \pi$~\cite{f1ref} and $b_1 \pi$~\cite{b1ref}.
While this is encouraging, as this state is more line with what
is expected from flux-tube models and the lattice in terms of mass and
decay modes, the data are of relatively limited statistical accuracy.  In
fact the quality of the data is such that strong conclusions regarding
the evidence for this state cannot be made.

\begin{figure}[htbp]
\vspace{7.5cm} 
\includegraphics{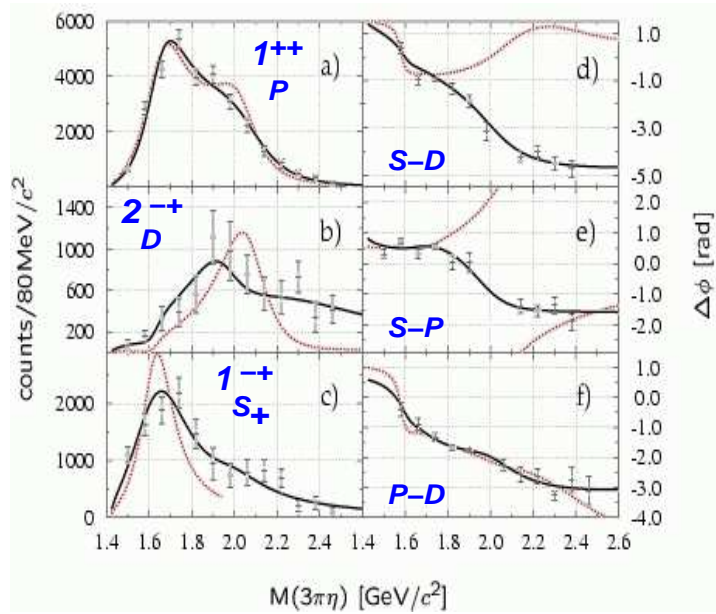}
\caption{PWA results for the $S$, $P$, and $D$ waves as a function
of the $f_1 \pi$ (or $(3\pi \eta)$) effective mass showing both the intensity 
and phase difference distributions~\cite{f1ref}.}
\label{pi2000_1_ss}
\end{figure}

\begin{figure}[htbp]
\vspace{8.2cm} 
\includegraphics{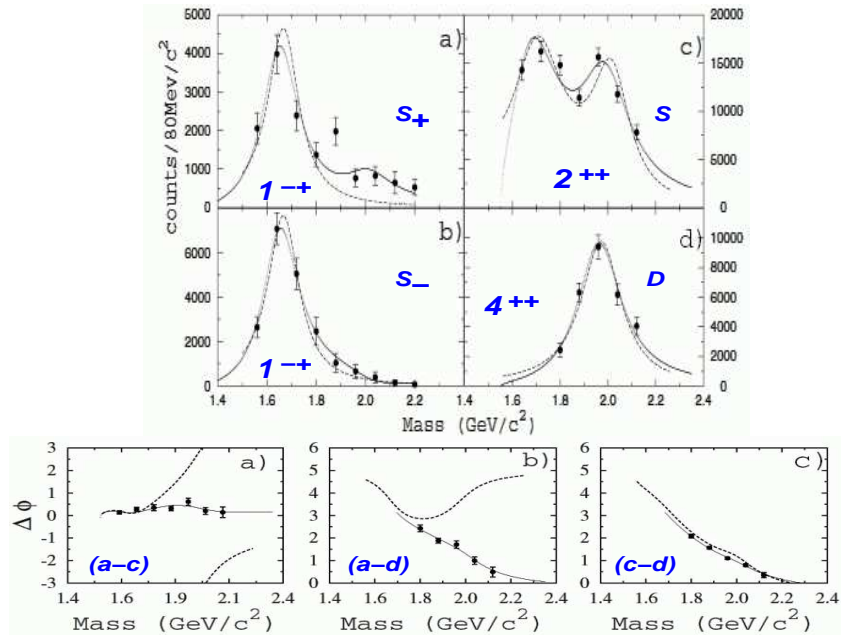}
\caption{PWA results for the $S$ and $D$ waves as a function
of the $b_1 \pi$ effective mass showing both the intensity and phase
difference distributions~\cite{b1ref}.  The phase difference plots in the
bottom row labeled $(a-c)$, $(a-d)$, and $(c-d)$, refer to the intensity
plots labeled as (a) through (d) in the top part of the figure.}
\label{pi2000_2_ss}
\end{figure}

The reported $\pi_1$(2000) from $f_1 \pi$ was seen through the reaction 
$\pi^- p \to \pi^- \pi^- \pi^+ \eta n$ (see Fig.~\ref{pi2000_1_ss}) and has 
a mass $M$=2001$\pm$30~MeV and width $\Gamma$=333$\pm$52~MeV.  
As seen from $b_1 \pi$ through the reaction $\pi^- p \to \omega \pi^0 \pi^- p$ 
(see Fig.~\ref{pi2000_2_ss}), the mass and width are quoted as 
$M$=2014$\pm$20~MeV and $\Gamma$=230$\pm$32~MeV.

\section{The Next Generation Experiment}

The specifications for the next generation experiment in meson spectroscopy
are already quite clear.  The essential ingredients are that the detector
should be hermetic for both charged and neutral particle final states, with
excellent resolution and particle identification capability.  This is
essential for successful and unambiguous partial wave analysis.  The beam 
energy must be high enough to allow for sufficient phase space for the
production of the exotic states, however it should not be too high, otherwise 
the cross section for background processes will increase and overwhelm the 
signals that are being sought.  The partial wave analysis requires high 
statistics experiments with sensitivity to production cross sections at the 
sub-nanobarn level.  

In terms of beam properties, the next generation experiment should be carried 
out with linearly polarized photons.  Photon beams are expected to be 
particularly favorable for the production of exotic hybrids, as the photon 
sometimes behaves as a virtual vector meson.  When the flux tube in this 
$S$=1 system is excited, both ordinary and exotic $J^{PC}$ are possible.  
In contrast, for an $S$=0 probe (e.g. pions or kaons), the exotic combinations 
are not generated.  To date, almost all meson spectroscopy experiments in the 
light-quark sector have been done with incident pion, kaon, or proton probes.  
High flux photon beams of sufficient quality and energy have not been 
available, so there are virtually no data on the photoproduction of mesons 
with masses below 3~GeV.  Thus up to now, experimenters have not been able to 
search for exotic hybrid mesons precisely where they are expected.  Theoretical
calculations indicate that the photoproduction cross sections of light-quark 
exotic mesons should be comparable to those for conventional meson states in 
this energy range~\cite{swat}.

As mentioned above, the next generation experiment should employ a linearly
polarized photon beam.  The diffractive production of a conventional meson 
takes place via natural parity exchange ($J^P = O^+,1^-,2^+,...$) in the 
intermediate state, whereas exotic meson production takes place via unnatural 
parity exchange ($J^P = O^-,1^+,2^-,...$).  Experiments that take place with 
an unpolarized or a circularly polarized photon beam cannot distinguish 
between the naturality of the exchanged meson.  However with a longitudinally 
polarized beam, one can distinguish the naturality of the intermediate state 
by selection based on the angle the polarization vector makes with the 
hadronic production plane.  This capability will be essential in helping to 
isolate the exotic waves in the data analysis.

\section{The Design of the GlueX Experiment}

The current Jefferson Laboratory accelerator can deliver electron beams
of up to nearly 6~GeV at currents of $\sim$200~$\mu$A and polarization
up to $\sim$80\%.  The planned upgrade of this facility is designed to
increase the maximum electron beam energy to 12~GeV, to construct a new
experimental Hall, called Hall D, for the GlueX experiment, and to lead
to equipment enhancements in the three existing Halls A, B, and C. The 
current schedule has the 12-GeV facility on track to begin its physics 
program in 2011-2012.

\begin{figure}[htbp]
\vspace{7.3cm}
\includegraphics{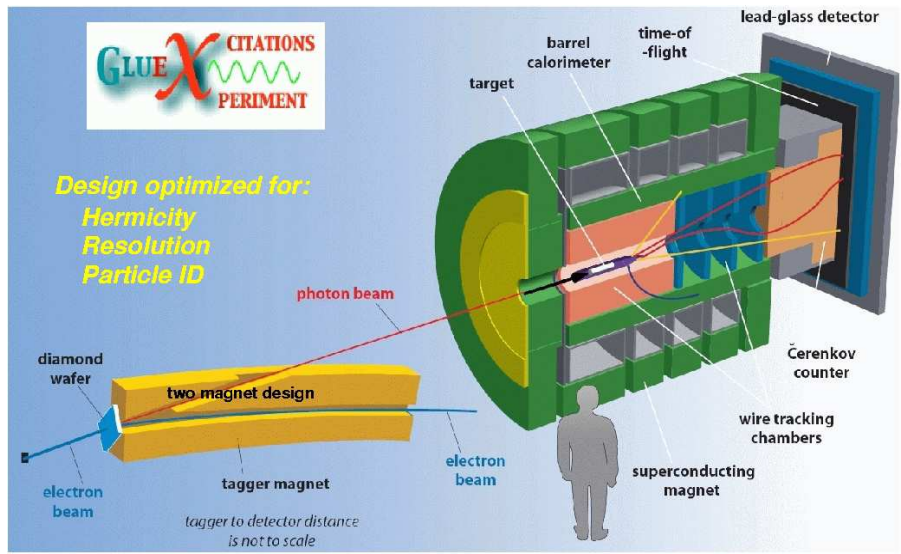}
\caption{Schematic picture of the GlueX detector and coherent bremsstrahlung 
photon tagging facility.  The subsystems within the solenoid are indicated.}
\label{gluex_det}
\end{figure}

The GlueX experiment in Hall D has been designed to perform meson
spectroscopy over a range of masses up to roughly 3~GeV~\cite{gluex}.  
Although GlueX aims to investigate the full spectrum of light conventional 
mesonic states, hybrid and exotic hybrid mesons, glueballs, multi-quark 
states, and molecular states, the exotic states will be the main initial 
thrust of the experimental program.  Fig.~\ref{gluex_det} shows a schematic 
layout of the planned hermetic detector.  The photon beam is incident on a 
liquid-hydrogen target that is located within a large 2~T superconducting 
solenoidal magnet.  The target is surrounded by drift chambers for charged 
particle tracking.  The tracking system includes a straw tube chamber for 
tracking at central and backward angles and a series of cathode strip 
chambers for tracking at forward angles.  Both of these packages will also 
provide for particle identification via $dE/dX$ measurements.  Surrounding 
the tracking detectors is a barrel calorimeter for particle identification 
and timing.  Filling the upstream end of the solenoid will be another 
calorimeter.  Downstream of the magnet is a large time-of-flight array, a 
{\v C}erenkov detector for particle identification, and a large lead-glass 
array that forms an electromagnetic calorimeter.  

The solenoidal geometry is ideally suited for a high-flux photon beam.  The 
electromagnetic charged particle background (electron-positron pairs) from 
interactions in the target is contained along the beam line by the axial 
field of the magnet.  The GlueX spectrometer has been designed for large and 
uniform acceptance for both charged and neutral particle final states, while 
providing the requisite resolution and particle identification capabilities 
for spectroscopy.

The GlueX photon beam will be produced using the coherent bremsstrahlung
technique.  Here the 12~GeV electron beam will impinge on a thin 
diamond crystal and will produce a linearly polarized beam 
of 9~GeV photons after collimation.  At the target the photon will achieve an 
average degree of linear polarization in the coherent peak of roughly 40\%.  
This detector system, along with the high duty factor of the electron beam 
from the accelerator make the search for hybrid mesons quite feasible.  
GlueX will collect enough data in its first year of operation with photon
fluxes of 10$^7$ to 10$^8$ s$^{-1}$ to exceed existing photoproduction data 
by several orders of magnitude.

Hand in hand with the design of the experimental equipment, the GlueX
Collaboration is performing detailed Monte Carlo studies of the physics, not
only to test and improve the design of the spectrometer, but also to develop 
the software tools essential for the success of the data analysis. The 
performance of the detector, the beam flux, and the linear polarization of the 
photon beam determine the level of sensitivity for mapping the hybrid spectrum.
The detector acceptance has been designed to be high and uniform in the
relevant meson decay angles in the Gottfried-Jackson frame for {\em both}
charged and neutral particle final states. A program of double-blind Monte
Carlo studies have been carried out using the GlueX partial wave analysis 
software.  These studies have shown that low-level exotic signals (at the
few percent level) can be successfully pulled out from the data, even with
relatively low statistics.  Even with extreme distortions in
these simulations, leakage effects due to the spectrometer are no more than
1\%.  To further certify the PWA codes, consistency checks with Monte Carlo
and real data will be made among different final states for the same decay
mode.

\section{Summary and Conclusions}

Understanding confinement requires an understanding of the glue that binds
quarks into hadrons.  Hybrid mesons are perhaps the most promising laboratory
to study the nature of the glue.  However, since their first observation,
their existence has been controversial, but a number of experimental results 
have provided tantalizing hints for the existence of these mesons.  Future 
studies, such as will be performed with the GlueX experiment at JLab, 
provide the hope for improved experimental results and interpretations.  
Here photoproduction promises to be rich in hybrids, starting with those 
having exotic quantum numbers where little or no data exist.  The GlueX 
experiment that will take place at the energy-upgraded Jefferson Laboratory, 
will employ photon beams of the necessary flux, duty factor, and polarization, 
along with an optimized state-of-the-art detector.  This experiment will 
provide for the detailed spectroscopy necessary to map out the hybrid meson 
spectrum, which is essential for an understanding of the confinement mechanism 
and the nature of the gluon in QCD.

\vskip 0.5cm
The author thanks the workshop organizers for the opportunity to discuss
the field of exotic meson spectroscopy.  He also gratefully acknowledges 
his fellow GlueX collaborators for their support and many efforts on the 
experiment to this point in time.  This work is supported by the 
National Science Foundation and the U.S. Department of Energy.

\end{document}